\documentclass{pasj00}
\usepackage[]{natbib}

\begin{document}
\SetRunningHead{Carlsson et al.}{High frequency acoustic waves}
\Received{2007/06/11}
\Accepted{}

\title{Can High Frequency Acoustic Waves Heat the Quiet Sun Chromosphere?}

\author{Mats \textsc{Carlsson}\altaffilmark{1}
        Viggo H.\textsc{Hansteen}\altaffilmark{1,2}
        Bart \textsc{De Pontieu}\altaffilmark{2}
        Scott \textsc{McIntosh}\altaffilmark{3,4}}
\email{Mats.Carlsson@astro.uio.no, Viggo.Hansteen@astro.uio.no, bdp@lmsal.com, mcintosh@boulder.swri.edu}
\author{Theodore D. \textsc{Tarbell}\altaffilmark{2}
        Dick \textsc{Shine}\altaffilmark{2}}
\email{tarbell@lmsal.com, shine@lmsal.com}
\author{Saku \textsc{Tsuneta}\altaffilmark{5}
        Yukio \textsc{Katsukawa}\altaffilmark{5}
        Kiyoshi \textsc{Ichimoto}\altaffilmark{5}
        Yoshinori \textsc{Suematsu}\altaffilmark{5}
        Toshifumi \textsc{Shimizu}\altaffilmark{6}
        Shin'ichi \textsc{Nagata}\altaffilmark{7}}
\email{saku.tsuneta@nao.ac.jp, yukio.katsukawa@nao.ac.jp, ichimoto@solar.mtk.nao.ac.jp, suematsu@solar.mtk.nao.ac.jp, shimizu@solar.isas.jaxa.jp, nagata@kwasan.kyoto-u.ac.jp}
\altaffiltext{1}{Institute of Theoretical Astrophysics, University of Oslo, PB 1029 Blindern, 0315 Oslo Norway}
\altaffiltext{2}{Lockheed Martin Solar and Astrophysics Laboratory, Palo Alto, CA 94304, USA}
\altaffiltext{3}{Department of Space Studies, Southwest Research Institute, 1050 Walnut St, Suite 400, Boulder, CO 80302, USA}
\altaffiltext{4}{High Altitude Observatory, National Center for Atmospheric Research, PO Box 3000, Boulder, CO 80307, USA}
\altaffiltext{5}{National Astronomical Observatory of Japan, Mitaka, Tokyo,
181-8588, Japan}
\altaffiltext{6}{ISAS/JAXA, Sagamihara, Kanagawa, 229-8510, Japan}
\altaffiltext{7}{Kwasan and Hida Observatories, Kyoto University,
Yamashina, Kyoto, 607-8471, Japan}

%


\maketitle

\begin{abstract}
We use Hinode/SOT Ca~{\sc ii} H-line and blue continuum broadband
observations to study the presence and power of high frequency
acoustic waves at high spatial resolution. We find that there
is no dominant power at small spatial scales; the integrated 
power using the full resolution of Hinode (0.05'' pixels, 0.16''
resolution) is larger than the power in the data degraded to 0.5'' pixels
(TRACE pixel size) by only a factor of 1.2. At 20~mHz
the ratio is 1.6. Combining this result with the
estimates of the acoustic flux based on TRACE data of
Fossum \& Carlsson (2006), we conclude that the total energy flux
in acoustic waves of frequency 5-40 mHz entering the
internetwork chromosphere of the quiet Sun is less than 800 W~m$^{-2}$,
inadequate to balance the radiative losses in a static
chromosphere by a factor of five.
\end{abstract}

\section{Introduction}

Observations show that the solar chromosphere radiates more than is
expected from radiative equilibrium. The extra energy needed to 
balance the radiative losses must be generated somewhere, transported
to the chromosphere and dissipated there. Acoustic waves were suggested
early on to play an important role \citep{Biermann1948,
Schwarzschild1948} because they are readily generated in the 
convection zone, can easily propagate and are expected to steepen
and dissipate in shocks in the chromosphere.

The estimates of the total radiative losses from the solar chromosphere
(and thus the required heating) are model dependent. \citet{Ulmschneider74}, \citet{Athay76} and
\citet{val3c} found the total loss to be 2500-3300 W m$^{-2}$,
2000-4000 W m$^{-2}$ and 4300 W m$^{-2}$, respectively. More recently,
\citet{Anderson+Athay1989} compared their computed model solar
chromospheres with the Vernazza, Avrett and Loeser models and found
that the VAL-C model is characterized by a total heat flux of 14000 W
m$^{-2}$, where about 90 \% is dissipated near the base of the
temperature plateau. The above estimates of radiative losses concerns
the average solar chromosphere. For the internetwork regions the
losses will be lower, the VAL3A model constructed to fit the lowest
intensities observed with Skylab has about 2.2 times lower radiative
losses than VAL3C \citep{Avrett1981}.

\citet{Carlsson+Stein1995,Carlsson+Stein1997} calculated synthetic
spectra of the strong chromospheric Ca\,II H-lines.
They showed that the enhanced chromospheric emission, which
corresponds to an outwardly increasing semi-empirical temperature
structure, can be produced by wave motion without any increase in the
mean gas temperature. However, these simulations do not reproduce
observations of spectral features formed in the middle to upper
chromosphere.  The reason could be the lack of waves above 20 mHz in
the simulations.  On the other hand, high-frequency acoustic waves,
even if produced in abundance in the convection zone, are heavily
radiatively damped in the photosphere \citep{Carlsson+Stein2002}: at
10 s period only 1\% of the generated acoustic energy flux remains at
a height of 500 km. In principle, waves with frequencies above what is
easily observed from the ground (above 20 mHz) and below what is very
heavily damped in the photosphere (below 50 mHz) could account for
the energy flux needed to balance the radiative losses from the
middle-upper chromosphere where the dynamic models fail. Such waves
are by many believed to constitute the dominant heating mechanism of
the chromosphere in non-magnetic regions.

The amount of acoustic energy contained in high frequency waves is
difficult to determine observationally for two reasons: First, image
distortions from the Earth's atmosphere introduces high frequency
noise in ground based observations. Second, the width of the intensity
formation region of any spectral diagnostic feature smears out the
signal of a short wavelength (high frequency) wave in both ground and
space based data. \citet{Fossum+Carlsson2005a} showed that the response
function of the 1600 {\AA} passband filter of the TRACE
\citep{TRACE} spacecraft is sufficiently narrow to allow the detection
of waves at least up to 40 mHz frequency. \citet{Fossum+Carlsson2005b}
used TRACE data to determine the acoustic energy flux of
waves in the 5-38 mHz range at the formation height of the 1600 {\AA}
passband integrated intensity and found a value of 0.44
kW~m$^{-2}$. A specially optimized observing sequence for
TRACE permitted a more detailed study of the acoustic wave
power as function of frequency \citep{Fossum+Carlsson2006}: a value of
0.51 kW~m$^{-2}$ in the 5-50 mHz range was found 
(note that the originally published value of 0.255 kW~m$^{-2}$ is too low by a factor
of two because of an error in a reduction routine). 
This value is too
low by a factor of 4-10 to balance the radiative losses of the
internetwork chromosphere as deduced from static models
\citep{Avrett1981, Anderson+Athay1989}.  As pointed out in their
paper, ``The major uncertainty in the analysis is the possibility of
high frequency power with spatial scales smaller than the resolution
element of TRACE.''  \citet{Cuntz+Rammacher+Musielak2007}
argues that, indeed, the high frequency power at small scales is
sufficient to make acoustic heating of the solar chromosphere locally
dominant. They base this conclusion on 
theoretical calculations of wave power generation, on 1D simulations of
acoustic waves (where high frequency waves can account for a temperature
increasing with height but where other aspects of the simulation that do
not agree with observations have to be attributed to the 1D limitation of
the simulation) and on synthetic images from 3D simulations where they
refer to the work of \citet{Wedemeyer-Bohm+Steiner+Bruls+Ramacher2007}.  
It is obvious that
high resolution {\em observations} would go a long way to answering the
question whether a significant acoustic energy flux is hidden in
spatial scales smaller than the resolution element of TRACE.

We here present such observations obtained with the Hinode
spacecraft at 0.16'' spatial
resolution (pixel size 0.05'').  In Section~2 we present the
observations and data reduction, in Sect.~3 we discuss the power
spectra of the observations at various spatial scales and we conclude
in Sect.~4.

\section{Observations and Data Reduction}\label{sec:observations}

\subsection{Choice of filters and response functions}

The TRACE 1600 passband is rather ideal for studying the
input of acoustic energy flux into the chromosphere because the
response of the passband integrated intensity to perturbations in
temperature is not very wide with a weighted mean formation height of
430~km \citep{Fossum+Carlsson2005a}, placing the sensitivity right at
the lower boundary of the chromosphere. As pointed out above, the
drawback is the large pixel-size of 0.5''. With the Solar Optical
Telescope (SOT) \citep{Tsuneta+etal2007} of Hinode \citep{Kosugi+etal2007} we have a
much higher spatial resolution, the diffraction limit ($\lambda\over
D$) is 0.16'' with 0.054'' pixel size, but can we find a filter that
is sensitive to temperature fluctuations at the lower end of the
chromosphere? The prime candidate is the filter centered at the
longward of the two resonance lines of singly ionized calcium --- the
Ca~{\sc ii}~H-line.  This filter is rather wide with an almost Gaussian
shape with FWHM of 0.22~nm, so the signal may be dominated by the
photospheric wings of the strong absorption line.  To investigate the
suitability of using Ca~{\sc ii}~H-line filtergrams for studies of
chromospheric dynamics we have determined the intensity response
function for the Hinode filter to perturbations of the
temperature. We follow the procedure of
\citet{Fossum+Carlsson2005a}. In general, a response function measures
the response of an observed quantity to a given
perturbation~\citep{Magain1986}. In this case the response function,
$R_{I,T}(h)$, measures the response of the relative change in intensity,
$\frac{\Delta I} {I}$, given a
perturbation in the temperature, $\Delta T(h)$, as function of
height in the atmosphere, $h$. The function is defined from
\begin{equation}
\frac{\Delta I}{I}=\int_{-\infty}^{\infty}R_{I,T}(h)\frac{\Delta
  T(h)}{T}dh .
\end{equation}
$R_{I,T}(h)$ is derived numerically from using a step function to
introduce a change of 1 \% in the temperature of a given reference
atmospheric model up to a given point in the atmosphere and varying
this point. For each such perturbation, the full non-LTE equations are
solved for a six level model atom of Ca~{\sc ii} with the code MULTI
\citep{Carlsson1986}, the resulting intensity profile of the H-line
was multiplied with the Hinode filter transmission profile
and integrated and the resulting Hinode intensities were
derivated with respect to depth giving the response function. It is
obvious that this approach has its limitations --- the response is
assumed to be linear and we may get a significant contribution from
heights where the acoustic waves have actually shocked, violating the
assumption of linearity. Furthermore, the derived response function
depends on the reference model atmosphere. Nevertheless, such a
response function give an indication as to what heights can be
diagnosed by intensity variations observed in a given filter.

\begin{figure}
  \begin{center}
    \FigureFile(85mm,65mm){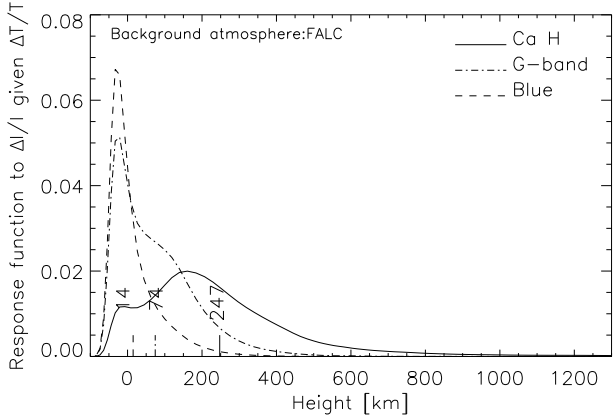}
  \end{center}
  \caption{Response functions for the SOT/BFI filters Ca~{\sc ii} H
({\it solid}), G-band ({\it dot-dashed}) and blue-continuum ({\it dashed})
using the FALC model of the average quiet solar atmosphere
as reference model atmosphere.}\label{fig:response_bc_cah_falc}
\end{figure}

Figure~\ref{fig:response_bc_cah_falc} shows the deduced temperature
response function for the SOT/BFI filters Ca~{\sc ii} H (centered at
396.86~nm with FWHM of 0.22~nm), G-band (centered at 430.64~nm with
FWHM of 0.63~nm), blue-continuum (centered at 450.51~nm with FWHM of
0.23~nm) (all wavelengths in air) using the FALC model of the average
quiet solar atmosphere \citep{Fontenla+Avrett+Loeser1990} as reference
model atmosphere. It is clear that the Ca~{\sc ii} H filter samples
the highest regions (average response height of 247~km) while the
G-band filter has a photospheric response with a upper photospheric
tail (because of the numerous spectral lines in the passband,
primarily from CH) with an average response height of 74~km and the
blue continuum has a clean photospheric response with an average
response height of 14~km.  
The response function of the Ca~{\sc ii} H
filter is particularly wide with a long tail extending into the middle
chromosphere. The width of the response function will tend to attenuate
the response to a wave perturbation --- more so for high frequency waves
with smaller wavelength. The attenuation as function of frequency of
the waves is given by the Fourier transform of the response function
\citep{Fossum+Carlsson2006}. For the Hinode calcium filter and the FALC
atmosphere we expect very little power beyond 20 mHz (only 5\% of the
amplitude remains). Note, however, that small scale vertical inhomogeneities 
would substantially increase the sensitivity to high frequency waves.
The blue continuum gives the cleanest photospheric response and has a
sharp response function.
For this study we therefore choose to observe with high cadence in the
calcium filter and the blue continuum filter.

\subsection{Observations}

The dataset analyzed in detail here was obtained on March 3 2007 between
05:48:03 and 07:09:29 UT and consists of 765 image pairs in the Ca~{\sc
  ii} H filter and the blue continuum filter. The cadence is strictly
fixed at 6.4~s with 3.2~s between the blue continuum and the calcium
image. We read out only the central 1024x512 pixels of the detector
(to keep the high cadence within the telemetry restrictions) thus
covering an area of 55''x28'' at Sun center. Exposure times were
0.41~s for the calcium images and 0.1~s for the blue continuum. After
736 frames there was a shift in the spacecraft pointing of 12'' and we
thus restrict the analysis to the first 736 frames in each channel,
covering a timespan of 1$^h$~18$^m$.

\subsection{Data Reductions}


We correct the data for dark current and camera artifacts using the
IDL routine \texttt{fg\_prep}, which is part of the Hinode
tree of solarsoft \citep{Tarbell+etal2007}. We also need to correct
for CCD sensitivity variations by flat-fielding the data. At the time of
analysis, there were no flat-field data in the Hinode tree of
solarsoft and we therefore constructed our own flat-fields. This was
done by extracting all the synoptic observations from disk-center in
the period November 2006 to April 2007. Frames that were closer in
time than 10 minutes were discarded as were frames containing pores or
sunspots. The flat-field was constructed from a mean of the remaining
440 frames. The calcium images have a solar rms variation of 16\%. The
averaging of 440 independent frames brings this down to an rms
variation of solar origin of 0.8\%, much smaller than the rms
variation of the flat-field image of 3\%.

\begin{figure}
  \begin{center}
    \FigureFile(85mm,65mm){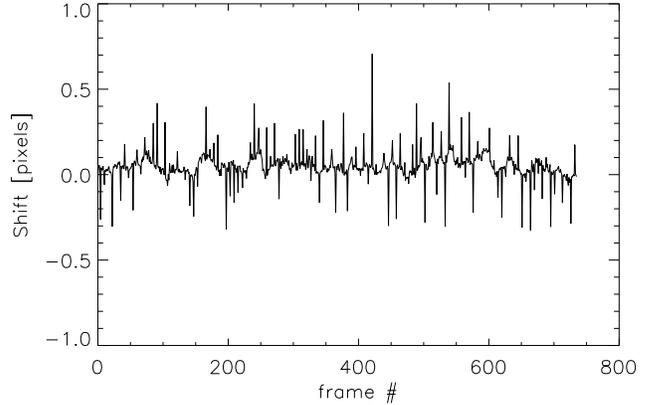}
  \end{center}
  \caption{Residual image jitter after the correlation tracker corrections,
calculated from cross-correlation frame to frame.}
\label{fig:alignment}
\end{figure}

The SOT correlation tracker removes most of the jitter introduced by
the spacecraft but visual inspection of the timeseries reveals there
is some remaining jitter. This is removed by performing a
rigid-co-alignment using cross-correlation from image to image and
applying the cumulative offsets to the whole
timeseries. Figure~\ref{fig:alignment} shows the deduced image to
image shift for the time-series showing that the residual image motion
after the correlation tracker corrections is on the order of 0.1
pixels or 0.005'' (one standard-deviation). The larger jumps are
caused by the replacement of the reference frame of the correlation
tracker approximately every 40~s \citep{Shimizu+etal2007}. 
The shifts calculated from
the calcium images and those calculated from the blue continuum images
are highly correlated. From this we conclude that the deduced rms is
actually uncorrected image motion and not errors in the
cross-correlation method. There is drift over the whole timeseries of
about 30 pixels or 1.7''. This means that there is an area close to
the edge of the field of view where we don't have data for the whole
timeseries.  This area is excluded from the further analysis.

\section{Power Spectra}\label{sec:power}

The distribution of power is known to be different in network areas
and in internetwork regions. The current dataset is very quiet and
does not contain any plage or pore but two small network patches. The
area of these patches is so small that the mean power is not affected
whether they are included or not. To facilitate comparisons with various
degrees of rebinning, we include the whole area in the following. For
each point we calculate the power of the intensity fluctuations
relative to the mean intensity
($\Delta I/<I>$) and take the mean. To determine the power as function
of spatial resolution, we redo the exercise several times after
rebinning the data to 2x2, 4x4, 8x8 and 16x16 original pixels per new
pixel.

\begin{figure}
  \begin{center}
    \FigureFile(85mm,57mm){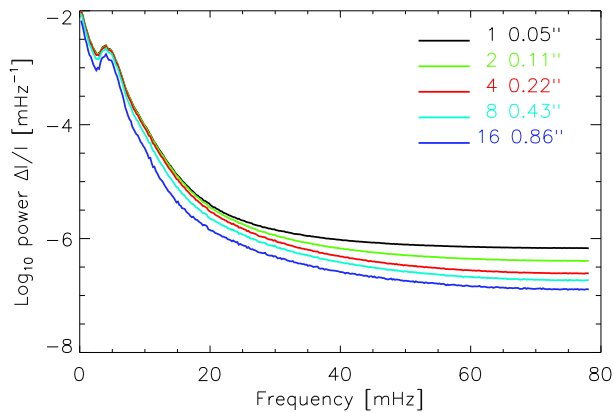}
  \end{center}
  \caption{Power of intensity fluctuations as function of frequency for
the SOT/BFI Ca~{\sc ii} H filter timeseries at various spatial binnings
(colour code in the legend).}
\label{fig:power_rebin_p129_cah}
\end{figure}

Figure~\ref{fig:power_rebin_p129_cah} shows the one-sided power (the
power at negative and positive frequencies added together) for the
calcium timeseries. Significant power can be seen up to about 30
mHz. The maximum power is at low frequencies, decreasing to 2.5 mHz
and then increasing through a local maximum at 4-5 mHz before
decreasing again. The noise level goes down with rebinning but very
little power is lost by the rebinning at 5-20 mHz: The integrated
power in the 5-20 mHz range is 90\% of the original value for the data
binned to 0.43'' pixels and 58\% for the data binned to 0.86''.

\begin{figure}
  \begin{center}
    \FigureFile(85mm,125mm){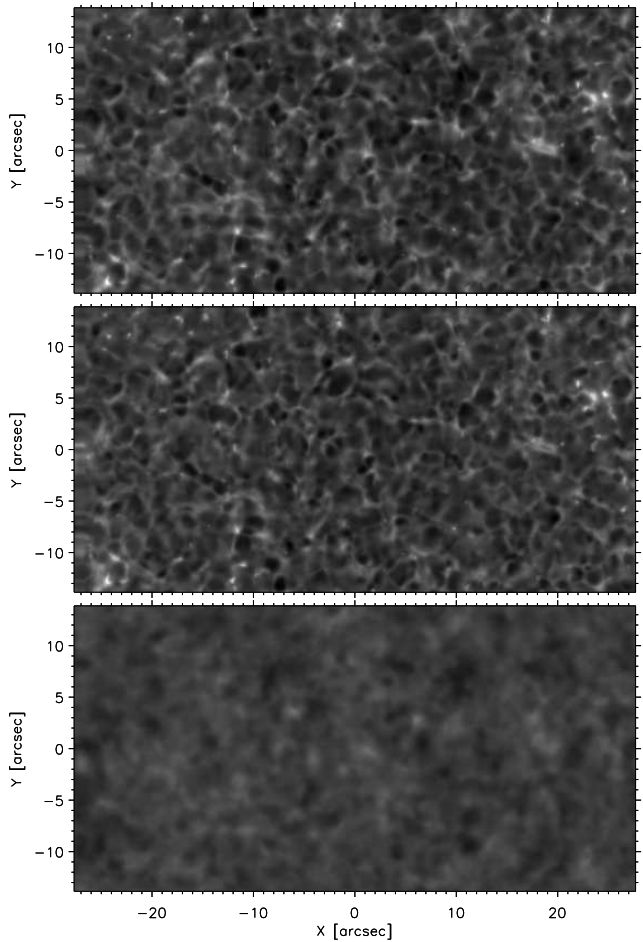}
  \end{center}
  \caption{Ca~{\sc ii} H filter image at 06:32:44 UT ({\it top}),
same snapshot with features moving slower ({\it middle}) and
faster ({\it bottom}) than 7~km~s$^{-1}$.
}
\label{fig:filter_images_p129_cah}
\end{figure}

\begin{figure}
  \begin{center}
    \FigureFile(85mm,125mm){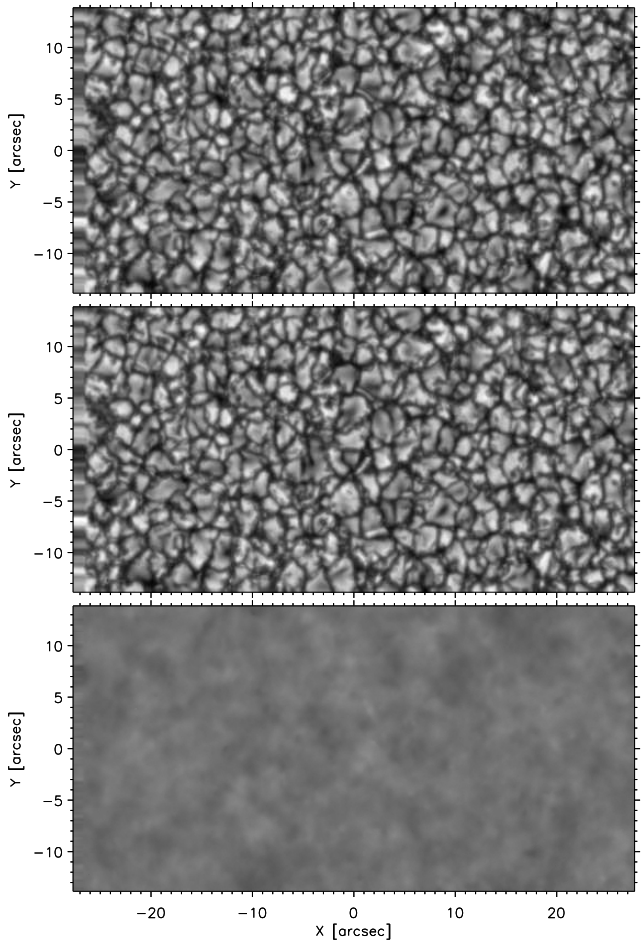}
  \end{center}
  \caption{Blue continuum filter image at 06:32:40 UT ({\it top}),
same snapshot with features moving slower ({\it middle}) and
faster ({\it bottom}) than 7~km~s$^{-1}$.
}
\label{fig:filter_images_p129_bc}
\end{figure}

The largest power is at the lowest frequencies. This is probably
caused by the granular evolution rather than by waves. We have tried
to separate the two components by performing a filtering of the
timeseries in the space-time domain with a conical filter. We
separate all components that have a horizontal velocity
smaller than 7 km~s$^{-1}$ (the approximate sound speed) from those
with a larger velocity. Sound waves that propagate horizontally should
have a phase speed of the intensity signal equal to the sound
speed. For an inclined wave, the phase speed will be larger (infinite
for a vertically propagating wave). All acoustic waves should thus be
contained in the high-pass part of the filtered data while granular
evolution (and gravity waves) will be in the low-pass
part. Figure~\ref{fig:filter_images_p129_cah} shows the original data
and the two filtered components for one snapshot. The original data
have some intensity modulations of large spatial scale that disappear in
the low-pass filtered image (e.g.\ a dark patch at [10,7] and a bright
patch at [2,0]). We find these patches in the high-pass component. It is
also clear that the smallest spatial scales are in the low-pass
component. Figure~\ref{fig:filter_images_p129_bc} shows the same for
the blue continuum filter.

\begin{figure}
  \begin{center}
    \FigureFile(85mm,125mm){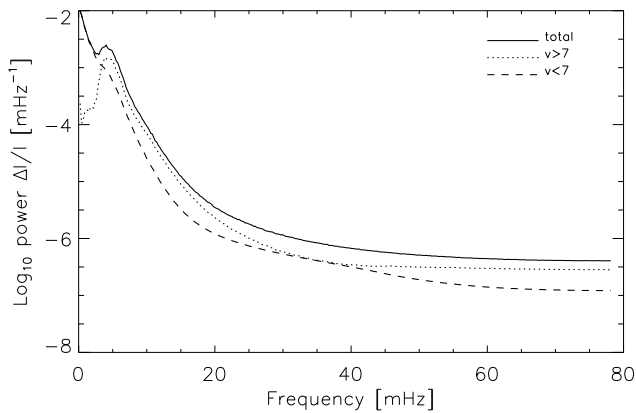}
  \end{center}
  \caption{Power of intensity fluctuations as function of frequency for
the SOT/BFI Ca~{\sc ii} H filter timeseries ({\it solid}), for the
low-pass filtered component (features moving slower than 7 km~s$^{-1}$)
({\it dashed}) and for the high-pass filtered component (features moving
faster than 7 km~s$^{-1}$) ({\it dotted}).
}
\label{fig:power_conefilt_p129_cah}
\end{figure}

Figure~\ref{fig:power_conefilt_p129_cah}
shows the power in the original data and the two filtered components.
The high power at low frequencies is totally dominated by the low-pass
component (evolution and gravity waves) while the acoustic wave component
gives rise to the peak at 5~mHz and dominates up to 20~mHz. The two 
components are of equal importance at 30~mHz while the noise is dominated
by the high-pass component (where spatially uncorrelated photon counting
noise is).

\begin{figure}
  \begin{center}
    \FigureFile(85mm,57mm){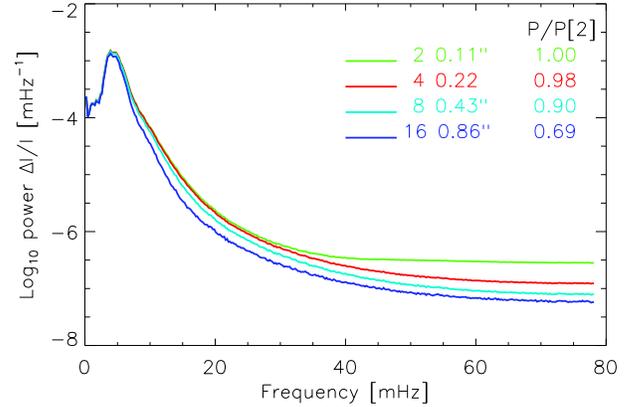}
  \end{center}
  \caption{As Fig.\ref{fig:power_rebin_p129_cah} but with the data first
filtered to remove features that move more slowly than 7 km/s horizontally.
The integrated power in the 5-20 mHz range compared with the unbinned data
is also given.}
\label{fig:power_rebin_vgt7_p129_cah}
\end{figure}

The power of the high-pass filtered data (where we expect the acoustic
waves to be) is shown in Fig.~\ref{fig:power_rebin_vgt7_p129_cah} which
also shows the integrated power in the 5-20 mHz range compared with
the 2x2 binned data (we compare with the 2x2 binned data because the
unbinned data has high velocity components from alignment
interpolations that cancel out when the data is first binned to
2x2. Furthermore, the 2x2 binned data has identical power spectrum as
the original data apart from the noise level, see
Fig.~\ref{fig:power_rebin_p129_cah}). The power ratios are similar to
the case for the unfiltered data except that the high-pass power at
low frequencies is identical for the various binnings, showing that
the differences at low frequencies in
Fig.~\ref{fig:power_rebin_p129_cah} are caused by slow
motion/evolution of sharp features. The integrated power 5-20~mHz is
90\% of the 2x2 binned power at 0.43'' binning and still 69\% at
0.86'' binning. The high percentages are because the integrated power
is dominated by the low frequencies close to 5~mHz where the
difference between the different binnings is very small. The acoustic
power will have a slower decline with frequency because the intensity
signal is depressed at higher frequencies from the significant width
of the response function \citep{Fossum+Carlsson2006}.  One would thus
expect that the small spatial scales are more important for the
integrated acoustic power than for the integrated intensity power.  As
an extreme case one can use the ratio of the intensity power at 20~mHz
as the ratio for the integrated acoustic power. This ratio is 69\% at
0.43'' binning and 43\% at 0.86'' binning.

\begin{figure}
  \begin{center}
    \FigureFile(85mm,57mm){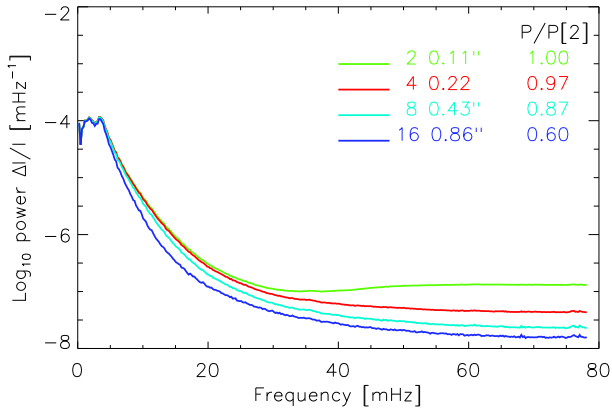}
  \end{center}
  \caption{As Fig.\ref{fig:power_rebin_vgt7_p129_cah} but for the blue
continuum filter.
}
\label{fig:power_rebin_vgt7_p129_bc}
\end{figure}

The Ca~{\sc ii} H-line is a scattering line with a broad response
function.  It could be argued that the spatial scales in the intensity
response of this line will be larger than the actual acoustic waves
scales and larger than for the TRACE 1600 filter. We have
therefore repeated the exercise for the blue continuum that has a much
more narrow response function.  The response is from the photosphere
so we expect the waves to have a much lower amplitude than in the
calcium filter. Figure~\ref{fig:power_rebin_vgt7_p129_bc} shows this to be
the case. The power is lower than for the calcium filter intensities
and there is a distinct 3~mHz peak from p-mode oscillations.  However,
the spatial scales of the high frequency signal is very similar in the
two filters: for the blue continuum the integrated power in the 5-20
mHz band is 87\% for 0.43'' binning and 60\% at 0.86'' binning. At
20~mHz the corresponding numbers are 65\% and 40\%, respectively.

For a proper comparison with TRACE, our data should first be
smeared with the PSF of TRACE and then rebinned to 0.5''
pixel size. The PSF is poorly known but is probably wider than the
diffraction limit. The aperture at the 1600 band of TRACE is
set by the 11.2~cm diameter of the filter. This corresponds to a
diffraction limit of 0.3''. TRACE is thus severly
undersampled and the PSF would have to be much broader than the
diffraction limit to add smearing beyond the pixel sampling. We thus
feel that the proper comparison is with the Hinode data
rebinned to 0.5''. As the numbers above show, even
rebinning to 0.86'' pixel size does not change the integrated mean power
of the timeseries much.

\section{Conclusions}\label{sec:discussion}

We have calculated the power of intensity variations as observed with
Hinode SOT/BFI Ca~{\sc ii} H and blue continuum filters at
high, strictly regular, cadence. We find that intensity power
corresponding to propagating acoustic waves is largest close to the
acoustic cutoff frequency. Furthermore, the spatial scales of the intensity
variations are predominantly larger than the
TRACE pixel size of 0.5'':  The integrated intensity power in
the 5-20 mHz range is 83\% of the original power when the data is
rebinned to the TRACE pixel size. Even at 20~mHz the power of
the rebinned data is 64\% of the original power. There is no large,
unseen, pool of high frequency waves at small spatial scales.
Combining this result with the work of
\citet{Fossum+Carlsson2005b,Fossum+Carlsson2006} we conclude that the
total energy flux in acoustic waves of frequency 5-40 mHz entering the
internetwork chromosphere of the quiet Sun is less than 800
W~m$^{-2}$, inadequate to balance the radiative losses in a static
chromosphere by a factor of five.

%

{\em We are grateful to the Hinode team for their efforts in the
  design, building and operation of the mission. Hinode is a Japanese
  mission developed and launched by ISAS/JAXA, with NAOJ as domestic
  partner and NASA and STFC (UK) as international partners. It is
  operated by these agencies in co-operation with ESA and NSC
  (Norway). SOT was developed jointly by NAOJ, LMSAL, ISAS/JAXA, NASA,
  HAO and MELCO. This work was supported by the Research Council of
  Norway grants 170926/V30 and 170935/V30.  B.D.P.  was supported by
  NASA contracts NNG06GG79G, NNG04-GC08G, NAS5-38099 (TRACE) and
  NNM07AA01C (HINODE).  SWM was supported by grants from the NSF
  (ATM-0541567) and NASA (NNG05GM75G, NNG06GC89G).}

\bibliographystyle{aa}

\end{document}